\documentclass[prl,aps,twocolumn,floatfix,showpacs]{revtex4}
\usepackage{graphics}
\usepackage{epsfig}
\usepackage{color}
\begin{document}

%\title{Bosons with long range dipolar interaction}
\title{Hexatic, Wigner Crystal, and Superfluid Phases of Dipolar Bosons}
\author{Kaushik Mitra, C. J. Williams and  C. A. R. S{\'a} de Melo}
\affiliation{Joint Quantum Institute, University of Maryland, College Park, Maryland 20742,\\ 
and National Institute of Standards and Technology, Gaithersburg, Maryland 20899}
\date{\today}

\begin{abstract}
The finite temperature phase diagram of two-dimensional dipolar bosons 
versus dipolar interaction strength is discussed. 
We identify the stable phases as dipolar superfluid (DSF), dipolar Wigner crystal (DWC),
dipolar hexatic fluid (DHF), and dipolar normal fluid (DNF). We also show 
that other interesting phases like dipolar supersolid (DSS) and dipolar hexatic 
superfluid (DHSF) are at least metastable, and can potentially be 
reached by thermal quenching.
In particular, for large densities or strong dipolar interactions, 
we find that the DWC exists at low temperatures, but melts into a
DHF at higher temperatures, where translational crystaline order
is destroyed but orientational order is preserved. Upon further increase in 
temperature the DHF phase melts into the DNF, where both 
orientational and translational lattice order are absent. 
Lastly, we discuss the static structure factor for some of the 
stable phases and show that they can be identified via 
optical Bragg scattering measurements.
 
\end{abstract}

\maketitle

%
%% Introduction
%

Arguably, one of the next frontiers in ultracold atomic and molecular physics is
the study of ultracold heteronuclear molecules such as  
KRb~\cite{mancini-2004,wang-2004,bongs-2006}, 
RbCs~\cite{kerman-2004}, and NaCs~\cite{haimberger-2004},
which can be produced using Feshbach resonances observed 
in mixtures of two types of alkali 
atoms~\cite{stan-2004,inouye-2004, ferlaino-2006}.
Thus, ultracold heteronuclear molecules consisting of Bose-Bose, 
Bose-Fermi, or Fermi-Fermi atom pairs offer many new opportunities 
because of their internal degrees of 
freedom~\cite{goral-2000,santos-2000,baranov-2002,rieger-2005,iskin-2007} 
such as their permanent electric dipole moment. 

The dipolar interaction between heteronuclear molecules is highly 
anisotropic in three dimensions having attractive and repulsive contributions. 
This makes it generally quite difficult to identify stable phases with 
good accuracy. In the particular case of bosonic heteronuclear 
molecules (Bose-Bose or Fermi-Fermi) the attractive part of the dipolar 
interaction may lead to undesired instabilities of the dipolar gas. 
However, the situation in two dimensions (2D) 
can be quite different, and arguably more interesting,
as the dipolar interaction can be made to be purely repulsive by the 
application of suitable static electric or microwave fields. 
In the case of bosonic dipolar molecules, 
several stable and metastable many body phases of 2D systems may be found.

In this manuscript, we obtain the finite and zero-temperature 
phase diagram of bosons interacting via short-range repulsive interactions 
$U$ and long-ranged dipolar interactions $E_D$ in two-dimensions. 
For weakly repulsive values of $U$, and small values of $E_D$ we find 
a dipolar superfluid phase of the Berezinskii-Kosterlitz-Thouless (BKT) type, 
which upon increasing values of $E_D$ becomes a dipolar Wigner crystal 
(DWC) forming a triangular lattice.
Numerical evidence for the existence of the DWC phase was obtained recently in 
Quantum Monte Carlo simulations of dipolar boson systems 
at zero temperature~\cite{buchler-2007,astrakharchik-2007}. 
Here, however, we develop an analytical variational theory that accounts 
not only for the superfluid to DWC phase transition at zero temperature, 
but also for the finite temperature melting of the DWC into a 
dipolar hexatic fluid (DHF), where crystalline translational order 
is destroyed but hexagonal orientational order is preserved. 
Further temperature increase leads to the melting of the hexatic phase 
into a dipolar normal fluid (DNF).
We also find that the dipolar supersolid phase (DSS), exhibiting 
both superfluid and crystalline order, has lower (higher) energy  
than the DSF (DWC) as density or dipole energy increases, but is 
at least metastable, thus being accessible using thermal quenching. 
The DSS phase can also melt into a dipolar hexatic superfluid (DHSF).
Lastly, we indicate that measurements of optical Bragg scattering  
can identify the DWC and DHF phases.

%
%% Hamiltonian
%
{\it Hamiltonian:} To describe the phases of interacting dipolar bosons in 2D, we start with 
the continuum Hamiltonian
\begin{equation}
\label{eqn:hamiltonian}
H  = - \frac{\hbar^2}{2m} \sum_i \nabla_i^2 +
 \frac{1}{2}\sum_{\langle i,j \rangle} \frac{D}{\vert {\bf r}_i - {\bf r}_j\vert^3} 
+ V_{loc}
\end{equation}
where $V_{loc} = \frac{1}{2} \sum_{\langle i,j \rangle} 
U\delta({\bf r}_i - {\bf r}_j) $, and the sum over $\langle i,j \rangle$
indicate the sum over all pairs of molecules. The first term of $H$ corresponds to 
the kinetic energy, the second to dipolar interactions, and the third $(V_{loc})$ 
to the local (short-range) interaction.

We begin our discussion of different ground states by analyzing 
first the dipolar Wigner crystal (DWC) phase. In this phase 
the dipolar interactions are dominant in comparison 
to the kinetic energy and local energy terms, 
such that the system crystalizes into a triangular lattice 
in two dimensions (2D). Thus, our variational wavefunction 
can be chosen to be  
\begin{equation}
\label{eqn:wigner-crystal-wavefunction}
\Psi_{\bf wc} ({\bf r}_1, {\bf r}_2, ...) = \frac{1}{\sqrt {\cal A}}
\sum_{p({\bf a})} \prod_i G_{ {\bf a}_i , \sigma } ( {\bf r}_i )
\end{equation}
%%
%
%$ \vert \psi \rangle$
where $\{{\bf a}_i\}$ forms a triangular lattice in 2D, and 
$A$ is the normalization constant. 
The $\sum_{p({\bf a}_i)}$ is a sum over all possible 
permutations $P$ of $\{{\bf a}\}$ and the function 
$G_{{\bf a}_i, \sigma } ( {\bf r}_i ) =
e^{-\left({\bf r}_i - {\bf a}_i\right)^2/2\sigma^2}/{\sqrt {\pi} \sigma}$ 
is a normalized Gaussian centered at the lattice site ${\bf a}_i$. 
We define the separation between neighboring lattice sites
to be $a$, and express the boson density as 
$\rho = 2/ (\sqrt{3}a^2)$. In addition, we introduce the
dimensionless parameters $r_D = 2m D\rho^{1/2}/\hbar^2$ 
($r_U = 2m U/\hbar^2$) as the ratio of the 
characteristic dipolar energy $E_D = D \rho^{3/2}$ 
(local energy $E_U = U \rho$) and kinetic energy 
$K = \hbar^2 \rho /2m$.

\begin{figure} [htb]
\centerline{ \scalebox{0.65} {\includegraphics{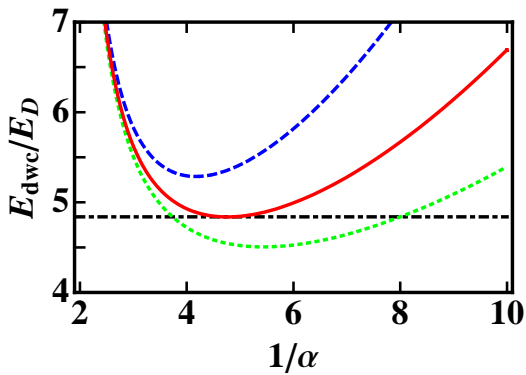}} } 
\caption{\label{fig:one}
(Color Online) Plots of the dipolar Wigner crystal energy $E_{\rm dwc}$ 
in units of the dipolar energy $E_D$ versus the inverse of the variational
parameter $\alpha$ for $r_D = 15$ (blue dashed line), 
$r_D = 26$ (red solid line), $r_D = 47$ (green dotted line). 
The horizontal (black dot-dashed line) is the energy $E_{\rm dsf}$ for a uniform 
superfluid.
}
\end{figure}

The variational wavefunction described in 
Eq.~(\ref{eqn:wigner-crystal-wavefunction}) is expected 
to be good for $r_D \gg 1$, where the dipole-dipole interaction 
is much larger than the kinetic energy, and can be used
to compute analytically the average kinetic, local and dipolar energies 
in terms of the variational parameter $\alpha = \sigma/a$, 
where $\sigma$ is the Gaussian width and $a$ is the lattice spacing. 
The dipolar Wigner crystal is only expected for $0 < \alpha < 1$,
where its average kinetic energy is 
%
%%
%%\begin{equation}
%%\label{eqn:kinetic-DWC}
$
\langle \Psi \vert K \vert \Psi \rangle_{\rm dwc} = 
N \sqrt3\hbar^2 \rho / ({4 m \alpha^2}).
$
%%\end{equation}
%%
%
The average local energy is 
%
%%
%%\begin{equation}
%%\label{eqn:local-DWC}
$
\langle \Psi \vert  V_{loc} \vert \Psi \rangle_{\rm dwc} = 
N  3 \sqrt3 U\rho P(\alpha) / ( {2 \pi \alpha^2}) 
$
%%\end{equation}
%%
%
where $P(\alpha) = 
\left( e^{-1/2 \alpha^2} + e^{-3 /2 \alpha^2} + e^{-2/ \alpha^2} \right)$
and the average dipolar energy is 
%
%%
%%\begin{equation}
%%\label{eqn:dipole-DWC}
$
\langle \Psi  \vert V_{dip} \vert \Psi \rangle_{\rm dwc} = 
N 3\sqrt{3} K D \rho \left[ P(\alpha)  +  F(\alpha)\right]/(2a) 
$
%%\end{equation}
%%
%
where $F(\alpha) = \left( 1 + 9\alpha^2/2 + 
225 \alpha^4/8 + 3675 \alpha^6/16 \right)$.
In general, $P(\alpha)$ and $F (\alpha)$ are represented by an infinite series 
in the variational parameter $\alpha$, but the series is rapidly convergent 
for $0 < \alpha < 1$, such that the first few terms are sufficient for 
the discussion of the total average energy 
$E_{\rm dwc} = \langle \Psi \vert H \vert \Psi \rangle_{\rm dwc}$.
The minimization of $E_{\rm dwc}$ with respect to 
$\alpha$ leads to the minima illustrated in Fig.~\ref{fig:one}, for $r_U = 0$
and a few values of $r_D$. For small values of $r_U$ the transition
is shifted towards lower values of $r_D$ (not shown in Fig.~\ref{fig:one}).

{\it Superfluid Phase:} The superfluid phase is more easily described by writing our original 
Hamiltonian in Eq.~(\ref{eqn:hamiltonian}) in second quantized notation, 
$H = K + V$, with the kinetic energy
%
%%
%%\begin{equation}
%%\label{eqn:kinetic-superfluid}
$
K = -\frac{\hbar^2}{2m}\int d{\bf r} \Phi^\dagger ({\bf r}) \nabla^2 \Phi({\bf r}) 
$
%%\end{equation}
%%
%
and potential energy
%
%%
%%\begin{equation}
%%\label{eqn:potential-superfluid}
$
V = \frac{1}{2} \int d {\bf r} d {\bf r'} \Phi^\dagger ({\bf r}) \Phi^\dagger ({\bf r'}) 
V ({\bf r}, {\bf r'})  \Phi ({\bf r}) \Phi ({\bf r'}),
$
%%\end{equation}
%%
%
where $ V ({\bf r}, {\bf r'}) =  D/\vert {\bf r} - {\bf r'}\vert^3 
+ U \delta({\bf r} - {\bf r'})$, and $\Phi^\dagger ({\bf r})$ 
is the bosonic field operator which creates a dipolar molecule 
at position ${\bf r}$. Describing the superfluid phase by the average 
$\langle \Phi^\dagger ({\bf r}) \rangle  =  
\langle \Phi ({\bf r}) \rangle = \sqrt \rho$
leads to the ground state energy
%
%%
%%\begin{equation}
%%\label{eqn:energy BEC}
$
E_{\rm dsf} = N \left[ \frac {3\pi \rho D}{2a} + \frac{U \rho}{2} \right], 
$
%%\end{equation}
%%
%
by assuming that the dipolar potential $V_{dip} (r) = D/r^3$ is unscreened 
for length scales $r \ge r_0$, and screened to $V_{dip} (r) = D/r_0^3$ for
length scales $r < r_0$, where $r_0 \approx 0.9 a$. 
%
%%This energy and phase coincide to that of 
%%the limit of $\alpha \to \infty$ of the dipolar Wigner Crystal phase.
%
In Fig.~\ref{fig:one}, the energy of the DSF phase is shown and compared
with that of the DWC phase.
%
%%a curve in the $r_D \times r_U$ plane showing
%%the phase boundary between the dipolar Wigner crystal and the superfluid phases 
%%at $T = 0$.
%
%

Next, we discuss the melting of the Wigner crystal phase, which occurs in two stages. 
First the dipolar Wigner crystal melts into a hexatic fluid, 
which does not have translational
order, but preserves rotational order. The melting occurs via the 
Kosterlitz-Thouless-Nelson-Halperin-Young (KTNHY) dislocation proliferation mechanism.
Second, the dipolar hexatic fluid transforms itself into 
the dipolar normal fluid by losing its rotational order
at a higher temperature.  

{\it Melting from Wigner Crystal to Hexatic Phase:} To study the melting of the DWC phase, 
we need to calculate its elastic energy. This calculation can be performed
by using a semiclassical approximation to the quantum-mechanical method for the calculation of 
elastic energies~\cite{nielsen-85}. The essential idea is to stretch the 
many-body wavefunction $\Psi ({\bf r}_1, {\bf r}_2, ...)$ on each particle
coordinate as $r_{i,\alpha} \to r_{i,\alpha} + \epsilon_{\alpha \beta}  r_{i \beta}$
(where repeated indices indicate summation), and expand up to second order in the
strain tensor $\epsilon_{\alpha \beta}$. The resulting
elastic energy is
\begin{equation}
\label{eqn:elastic-energy}
E_{\rm el} = \frac{1}{2} \int d{\bf r} \left[ 2 \mu \epsilon_{\alpha \beta}^2
+ \lambda \epsilon_{\alpha \alpha}^2 \right],
\end{equation}
where $\mu = 15\sqrt{3} D/4a^5$ and $\lambda  = 3\mu $ 
are the unrenormalized Lam{\'e} coefficients,
and the symmetric (rotation-free) strain tensor is 
$
\epsilon_{\alpha \beta} ({\bf r}) = \frac{1}{2} 
\left[ 
\frac{\partial u_\alpha ({\bf r})}{\partial {\bf r}_\beta}
+ \frac{\partial u_\beta ({\bf r})}{\partial {\bf r}_\alpha}
\right],
$
with $u_{\alpha} ({\bf r})$ being the displacement from equilibrium
position.
%
%% *** Check carefully the relation with Nelson's
%% work, specifically the elastic energy ***
%

We follow Ref.~\cite{nelson-79}, and 
decompose the strain tensor $\epsilon_{\alpha \beta} ({\bf r})$ 
into a regular (smoothly varying) $\phi_{\alpha \beta} ({\bf r})$ 
and a singular (dislocation) $d_{\alpha \beta} ({\bf r})$ contribution.
Using $k_B = 1$, the Hamiltonian for the smooth part can be written as 
\begin{equation}
\label{eqn:regular-hamiltonian}
\frac{H_{\rm reg}}{T} = \frac{1}{2} \int \frac{d{\bf r}}{{a}^2} 
\left[
2 {\bar \mu} \phi_{\alpha \beta}^2 + {\bar \lambda} \phi_{\alpha \alpha}^2
\right],
\end{equation}
where ${\bar \mu} = \mu a^2/T$ and ${\bar \lambda} = \lambda a^2/T$.
Additionally the Hamiltonian for the singular part can be written as
\begin{equation}
\label{eqn:singular-hamiltonian}
\frac{H_{\rm dis}}{T} = - \frac{K}{8 \pi} \sum_{{\bf r} \ne {\bf r'}}
b_{\alpha} ({\bf r}) \Gamma_{\alpha \beta} b_{\beta} ({\bf r'})
+ \frac{E_c}{T} \sum_{{\bf r}} \vert {\bf b} ({\bf r}) \vert^2,
\end{equation}
where $b_{\alpha}$ is the alpha-component of the Burger's vector ${\bf b}$
defined by the contour integral of the displacement field ${\bf u}$ around the
dislocation: $\int d {\bf u} = a {\bf b} ({\bf r})$. 
Also, the coefficient $K = K_0 a^2/T$, is related to the unrenormalized 
Lam{\'e} constants through $K_0 = 4 \mu (\mu + \lambda)/(2\mu + \lambda)$, 
and $E_c = (\eta + 1) K T/8\pi$ is the core energy associated with a dislocation
of core diameter $\eta a$, while the interaction coefficient is  
\begin{equation}
\label{eqn:gamma}
\Gamma_{\alpha \beta} = \delta_{\alpha \beta} 
\ln \left[ \frac{\vert {\bf r} - {\bf r'} \vert}{a} \right] - 
\frac{\left[{\bf r} - {\bf r'}\right]_\alpha \left[{\bf r} - {\bf r'}\right]_\beta}
{\vert {\bf r} - {\bf r'} \vert^2}
\end{equation}

To obtain the critical temperature for the melting $T_{m}$ of the dipolar Wigner crystal
we solve the renormalization flow equations for the elastic parameters 
\begin{eqnarray}
\label{eqn:lame-renormalized}
\frac{d\bar{\mu}^{-1}}{dl} &=& 
3\pi y^2(l)e^{K(l)/8\pi}  \nonumber\\
\frac{d\left[\bar{\mu}(l)+ \bar{\lambda}(l)\right]^{-1}}{dl} &=& 
3\pi y^2(l)e^{K(l)/8\pi}
\left( F_0(l) + F_1(l) \right) \nonumber,
\end{eqnarray}
of the Hamiltonian $H_{\rm reg}$,
where $F_n (l) = I_n \left(K(l) / 8\pi \right)$, with $I_n (x)$ being the
modified Bessel function of order $n$. These equations need to be solved 
self-consistently with the flow equations for the parameters 
\begin{eqnarray}
\label{eqn:dislocation-renormalized}
\frac{dy(l)}{dl} &=& 
\left( 2 - \frac{K(l)}{8\pi}\right)y(l) + 2\pi y^2(l)e^{K(l)/16\pi} F_0 (l) \nonumber\\
\frac{dK^{-1}}{dl} &=& 
\frac{3}{2}\pi y^2(l)e^{K(l)/8\pi}
F_0 (l) - \frac{3}{4}\pi y^2(l)e^{K(l)/8\pi}F_1 (l) \nonumber,
\end{eqnarray}
of the dislocation Hamiltonian $H_{\rm dis}$. All flow equations are accurate to
order $y^3$, where $y = e^{-E_c/T}$ plays the role of the fugacity.   
The critical temperature $T_h$ is reached when 
\begin{equation}
K (T_h) = \frac {4 \bar{\mu} (T_h) \left[ \bar{\mu} (T_h) + \bar{\lambda} (T_h) \right]}
{2 \bar{\mu} (T_h) + \bar{\lambda} (T_h)} 
= 16\pi
\end{equation}

\begin{figure} [htb]
\centerline{ \scalebox{0.65} {\includegraphics{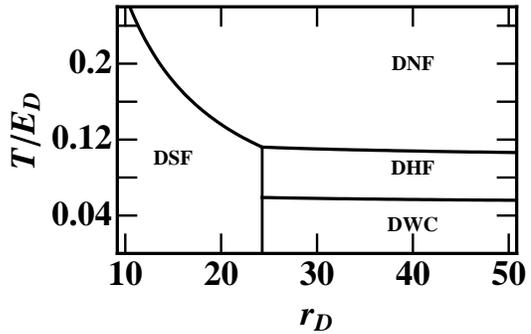}} } 
\caption{\label{fig:two}
Finite temperature phase diagram of $T/E_D$ versus $r_D$ showing the
dipolar superfluid (DSS), dipolar Wigner crystal (DWC), dipolar
hexatic (DHF) and dipolar normal fluid (DNF) phases. 
phases. 
}
\end{figure}

Solving the RG flow equations, we get the melting temperature from the
Wigner crystal to the hexatic phase to be $T_h = 0.05E_D$ in the 
classical limit of $r_D \to \infty$.
%
%% where quantum mechanical particle exchanges are not important 
%
The melting temperature $T_h$ separating the DWC and DHF phases is shown
in Fig.~\ref{fig:two}. 

{\it Transition from  Hexatic to Normal Phase:} 
As the dipolar Wigner crystal melts at $T_h$, translational order disappears
but orientational order is preserved, with the emergence of the hexatic
order parameter $\Psi_6 ({\bf r}) = e^{6i \theta ({\bf r})}$. The bond-angle
field $\theta ({\bf r})$ between the location of the center of masses of the
dipolar bosons (heteronuclear molecules) is related to the displacement field
$u({\bf r})$ by $\theta ({\bf r}) = \left[ \partial_x u_{y} ({\bf r})  - \partial_y 
u_{x} ({\bf r})\right]/2$. The elastic energy for such a situation can be also
obtained using a semiclassical approximation to the 
method of Ref.~\cite{nielsen-85}, where each particle coordinate
in the many-body wavefunction $ \Psi ( {\bf r}_1, {\bf r}_2, ... )$ is locally rotated 
$r_{i,\alpha} \to M_{\alpha \beta} r_{i,\beta}$, where $M_{\alpha \beta}$ is a local
rotation matrix (tensor). For the triangular lattice considered here the
elastic energy becomes
\begin{equation}
\label{eqn:hexatic-energy}
E_{\rm he} = \frac{\Gamma_6}{2}\int 
\frac{d \bf r}{a^2} \vert \nabla \theta ({\bf r}) \vert^2,
\end{equation}
where $\Gamma_6 \approx 2 E_c$ is the phase stiffness of the hexatic phase,
which is directly related to the dislocation core energy 
$E_c = 1.1 K T/8\pi$. 
%
%%The free energy for the hexatic phase described above has exactly 
%%the same form of the free of the celebrated 2D-XY 
%%model~\cite{kosterlitz-74}. 
%
The critical temperature for the disappearance of hexatic order 
and the emergence of the normal phase $T_n$ is then determined by 
the RG flow of the 2D-XY model~\cite{kosterlitz-74}, which leads
to the condition $\Gamma_6 (T_n) = 72 T_n/\pi$.  The algebraic decay of orientational
order is then destroyed at temperature $T_n$ by the dissociation of pairs
of $\pm \pi/3$ disclinations, which play the role of vortices and anti-vortices
of the standard 2D-XY model.
In Fig.~\ref{fig:two}, the critical temperature $T_n$ separating the phases
DHF and DNF is shown as a function of $r_D$.
When $r_D \to \infty$, we obtain $T_n \to 0.11E_D$ which is larger than the 
melting temperature of the dipolar Wigner crystal.

\begin{figure} [htb]
\centerline{ \scalebox{0.65} {\includegraphics{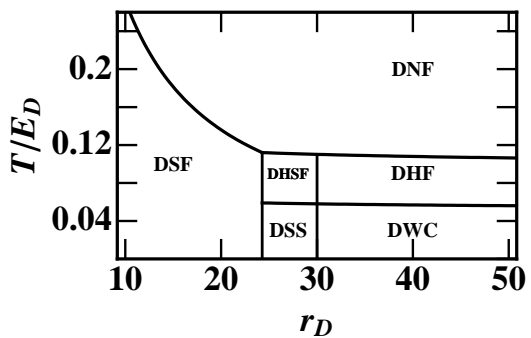}} } 
\caption{\label{fig:three}
Phase diagram showing possible dipolar supersolid (DSS) and dipolar
hexatic superfluid (DHSF) phases.
}
\end{figure}

{\it Supersolid Phase:} 
Starting from the Hamiltonian defined in the superfluid section 
we seek a variational solution
\begin{equation}
\label{eqn:supersolid-wavefunction}
\langle \Phi ({\bf r}) \rangle_{\rm ss} = 
{\widetilde \Phi}_{\rm ss} ( {\bf r} )
=  \sqrt{\rho_{\rm sf}} + 
\sqrt{\rho_{\rm ss}} \sum_i G_{{\bf a}_i , \sigma_{\rm ss}} ({\bf r}),
\end{equation}
for the supersolid phase, where the Gaussian functions 
$G_{{\bf a}_i , \sigma_{\rm ss}} ({\bf r})$ form a triangular 
lattice of side $a_{\rm ss}$. The normalization condition is 
$\int d{\bf r} \vert {\widetilde \Phi}_{\rm ss} ({\bf r}) \vert^2/V = \rho$,
while $\sigma_{\rm ss} = \beta a_{\rm ss}$ is the gaussian width and 
$\beta$ is the corresponding variational parameter.  
The supersolid order parameter 
${\widetilde \Phi}_{\rm ss} ({\bf r})$ describes a non-uniform superfluid 
with both off-diagonal long-range order due to broken $U(1)$ symmetry 
and diagonal long-range order due to the discrete lattice symmetry.
When $\rho_{\rm sf} \gg \rho_{\rm ss}$, the energy 
is essentially that of a superfluid, and when $\rho_{\rm sf} \ll \rho_{\rm ss}$, 
the average kinetic energy is
$K_{\rm dss} = N \sqrt3\hbar^2 \rho / ({4 m \beta^2})$,
while the total potential energy
is $V_{\rm dss} = V_1 + V_2$, 
where $V_1 = N 3 \sqrt3 U\rho  P(\beta) / ( {2 \pi \beta^2})  
+ N \sqrt{3} U \rho /8\pi \beta^2$ is the local potential energy 
and $V_2 = N 3\sqrt{3} K D \rho  \left[ P(\beta)  +  F(\beta)\right]/(2a_{\rm ss}) 
+ N \sqrt{3} D \rho /4a_{\rm ss}^3$ is the dipolar energy. 

We find that the total energy $E_{\rm dss} = K_{\rm dss} + V_{\rm dss}$ of the 
dipolar supersolid (DSS) phase can be lower than that of 
the superfluid phase due to the contributions 
coming from the dipolar interaction, but $E_{\rm dss}$ is always higher than 
the total energy $E_{\rm dwc}$ of the dipolar Wigner crystal for the same values 
of parameters. Thus, the transition from the superfluid to 
the supersolid phase does not occur within our variational 
ansatz, even if we include additional correlations via Jastrow 
factors~\cite{astrakharchik-2007}.  However, this transition can not
be excluded, because only a few classes of variational wavefunctions have been
explored.  Additionally, we find that the 
supersolid phase is at least metastable, since $E_{\rm ss}$ has a minimum 
that can be reached upon thermal quenching,
and possibly also melted into a metastable dipolar hexatic superfluid (DHSF).
A phase diagram showing possible DSS and DHSF phases is presented in 
Fig.~\ref{fig:three}.  
%
%% *** say something about the location of the minimum
%%(for instance does it happen before the WC minimum) ***
%

{\it Experimental characterization of the various phases:}
The various phases proposed here can be characterized by the measurement
of their density-density correlations, as reflected in the structure factor
$S({\bf q}) = \langle n({\bf q}) n ({\bf -q}) \rangle$,
where $n ({\bf q})$ is the Fourier transform of the density operator 
$n ({\bf r})$, and the symbol $\langle ... \rangle$ corresponds to 
both quantum mechanical and thermal
averaging.

\begin{figure} [htb]
\centering
\begin{tabular}{cc}
\epsfig{file=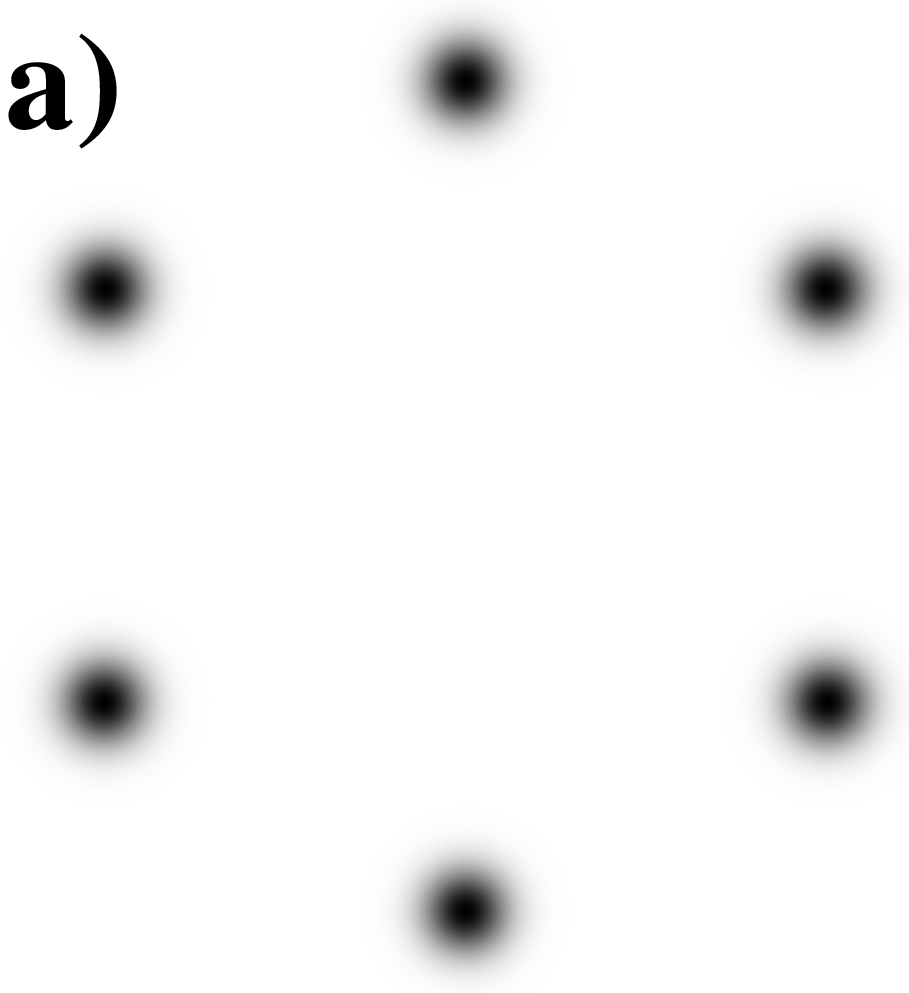,width=0.29\linewidth} \quad \quad \quad \quad &
\epsfig{file=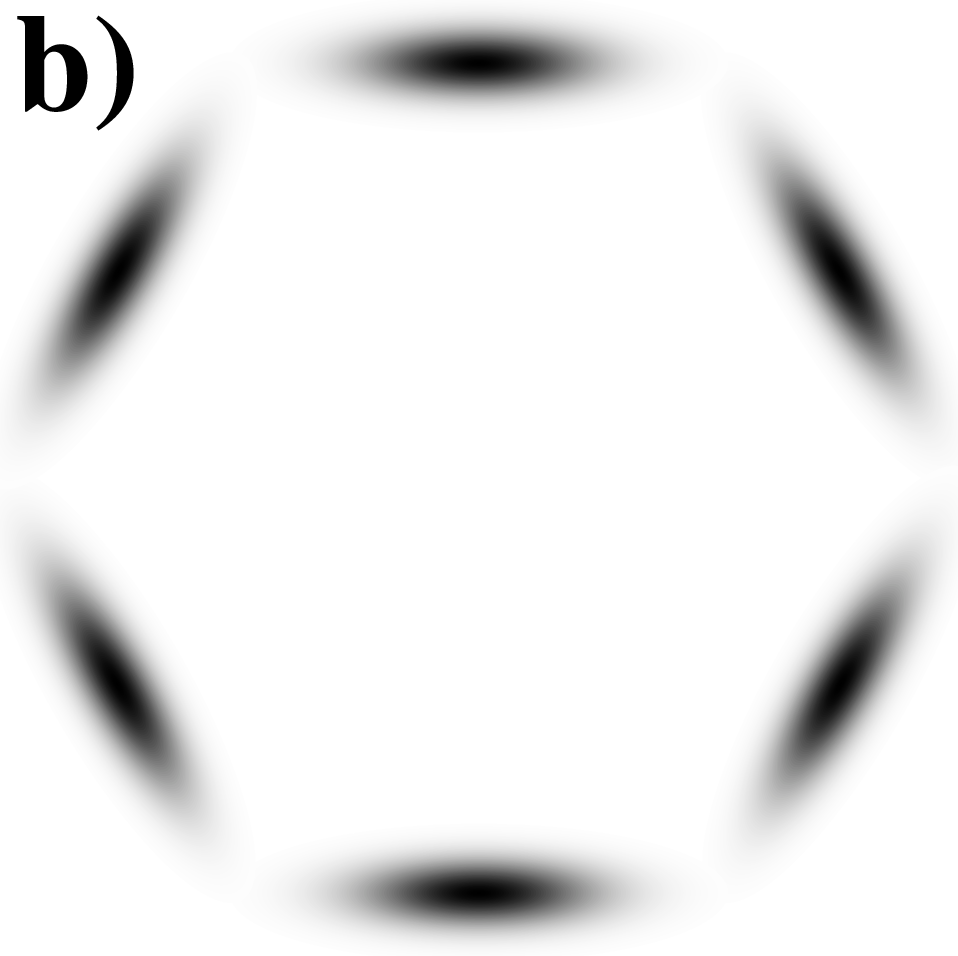,width=0.29\linewidth}  
\end{tabular}
\caption{\label{fig:four}
Bragg scattering patterns near the first reciprocal vectors
for  a) the dipolar Wigner crystal phase 
and for b) the dipolar hexatic fluid phase.
}
\end{figure}

The most dramatic effects in the structure factor are found 
in dipolar Wigner crystal phase where 
$S({\bf q}) \sim \vert {\bf q} - {\bf G} \vert^{-2 + \nu ({\bf G}, T)}$
revealing the power law behavior characteristic of two dimensions
in the vicinity of the reciprocal lattice (Bragg) vectors ${\bf G}$.
The first Bragg vector has magnitude $\vert {\bf G}_1 \vert = 4 \pi/\sqrt{3} a$,
and the Bragg scattering pattern then exhibits 6-fold symmetry  
below the melting temperature $T_m$ as shown in Fig.~\ref{fig:four}a.
The exponent $\nu ({\bf G}, T) = (\vert {\bf G}_1 \vert/ 4 \pi) (T/\mu_{*}) 
(3 \mu_{*} + \lambda_{*})/(2 \mu_{*} + \lambda_{*})$ is related to 
the renormalized Lam{\' e} coefficients $\mu_{*}$ and $\lambda_{*}$.
This reflects the power law decay of the correlation function
$C ( {\bf G}, {\bf R} ) \sim \vert {\bf R} \vert^{-\nu ({\bf G}, T)}$,
which is the Fourier transform of $S({\bf q})$. 
The profile of $S ({\bf q})$ for the dipolar hexatic phase corresponding 
to a melted dipolar Wigner crystal with orientational order is characterized 
by the hexagonal pattern shown in Fig.~\ref{fig:four}b. 

{\it Conclusions:} We have described some of the possible phases of dipolar bosons in two 
dimensions, including superfluid, supersolid, hexatic superfluid, Wigner crystal, 
hexatic and normal fluids. Within our variational approach we have 
concluded that the supersolid and hexatic superfluid phases may have lower 
free energy than the superfluid phase in some region 
of the phase diagram, but do not seem to have lower free energy than 
the Wigner crystal 
or hexatic fluid in the same region. 
%
%%Even when more correlated 
%%variational wavefunctions 
%%including Jastrow factors are used, these results seem to persist. 
%
However, the supersolid and hexatic superfluid phases are 
at least metastable, and thus may be reached via thermal quenching, 
and probed via Bragg scattering. Furthermore, we have shown that dipolar 
Wigner crystal does not melt directly into a normal fluid, but presents a 
two-stage melting first into a hexatic phase which preserves orientational order, 
and then into a normal fluid upon further increase in temperature.
Finally, we indicated the experimental signatures of stable phases with 
translational or orientational order in a Bragg spectroscopy measurement 
that detect the static structure factor.

\end{document}